# Edge effect removal in Fourier ptychographic microscopy via periodic plus smooth image decomposition

**An Pan,[a,b,*] Aiye Wang,[a,b,d] Junfu Zheng,[c] Yuting Gao,[a,b,d] Caiwen Ma,[a,b,d] Baoli Yao[a,b,*]**
[a] State Key Laboratory of Transient Optics and Photonics, Xi'an Institute of Optics and Precision Mechanics, Chinese Academy of Sciences, Xi'an 710119, China
[b] University of Chinese Academy of Sciences, Beijing 100049, China
[c] The Institute of Optics, University of Rochester, Rochester, New York 14627, USA
[d] CAS Key Laboratory of Space Precision Measurement Technology, Xi'an 710119, China

**Abstract**. Fourier ptychographic microscopy (FPM) is a promising computational imaging technique with high resolution, wide field-of-view (FOV) and quantitative phase recovery. So far, a series of system errors that may corrupt the image quality of FPM has been reported. However, an imperceptible artifact caused by edge effect caught our attention and may also degrade the precision of phase imaging in FPM with a cross-shape artifact in the Fourier space. We found that the precision of reconstructed phase at the same subregion depends on the different sizes of block processing as a result of different edge conditions, which limits the quantitative phase measurements via FPM. And this artifact is caused by the aperiodic image extension of fast Fourier transform (FFT). Herein, to remove the edge effect and improve the accuracy, two classes of opposite algorithms termed discrete cosine transform (DCT) and periodic plus smooth image decomposition (PPSID) were reported respectively and discussed systematically. Although both approaches can remove the artifacts in FPM and may be extended to other Fourier analysis techniques, PPSID-FPM has a comparable efficiency to conventional FPM algorithm. The PPSID-FPM algorithm improves the standard deviation of phase accuracy as a factor of 4 from 0.08 radians to 0.02 radians. Finally, we summarized and discussed all the reported system errors of FPM within a generalized model.

***An Pan, E-mail: panan@opt.cn, Baoli Yao, E-mail: yaobl@opt.ac.cn**

## 1 Introduction

Fourier ptychographic microscopy (FPM) invented in 2013 by Zheng and Yang et al. [1-3] is a promising computational imaging technique with high resolution (HR), wide field-of-view (FOV), and quantitative phase recovery, sharing its root with aperture synthesis [4-6] and phase retrieval [7-9]. The insight is that the objective can only collect light ranging a certain angle, characterized by numerical aperture (NA). However, parts of the scattering light with higher angle illumination can be also collected due to light matter interaction and the sample's high frequency information can be modulated into the passband of objective lens. Instead of conventionally stitching small HR image tiles into a large FOV, FPM uses a low NA objective to take advantage of its innate large

FOV and stitches together low resolution (LR) images in Fourier space to recover HR. Herein, no mechanical scanning and refocusing are required to achieve a complex gigapixel image. Currently, FPM is developed toward high-throughput imaging [10, 11], high-speed and single shot imaging [12-15], three-dimensional (3D) imaging [16-19], mixed-state decoupling [20, 21] and biomedical applications [22-26].

One of its successful applications is for quantitative phase imaging. Plenty of system calibration methods were reported for high-precision FPM, e.g., aberration removal [27], LED intensity fluctuation correction [28], LED position correction [29], vignetting effect removal [30], and noise suppression [31, 32], which has been summarized in Ref. 3 completely. However, we found an imperceptible artifact caused by edge effect that may degrade the imaging precision of FPM, especially the phase recovery, because of the aperiodic image extension via fast Fourier transform (FFT) and is seldom reported in FPM, while it has been found in the field of transport of intensity equation (TIE) [33, 34]. This artifact is a cross shape in Fourier space spanning from high frequency to low frequency. We found that the high frequency artifacts are corresponding to the obvious fluctuations at the edge of FOV in spatial domain, which is similar to Gibbs effect [35] or ringing effect, while the low frequency parts look like "ripples" at the background of reconstructions, especially obvious for those pure phase samples. Even worse, the imaging precision of the same subregion depends on the different sizes of block processing due to different edge conditions, which limits the quantitative use of FPM.

A straightforward method to alleviate the edge effect may be zero-padding interpolation, i.e., zeros are added to the edges of the image in the spatial domain to meet the continuity of four boundaries. Nevertheless, the edge effect exists still, because the edge effect is mainly caused by the aperiodicity of specimens when using FFT, and the images with zero-padding interpolation are still aperiodic. Hence, this simple interpolation method is ineffective.

In this paper, we reported two typical methods, termed discrete cosine transform (DCT) and periodic plus smooth image decomposition (PPSID) to eliminate the edge effect in FPM from two opposite perspective. The principle of DCT method is to “add information” into the original image to making the original image periodic by quadrupling the original image and remove the reduplicative information after FFT by cropping a quarter. Since the DCT method has been used in TIE [33, 34], we illustrated it in brief. While inspired from the mathematical work [36], the PPSID approach “reduce” the unnecessary information or artifact after FFT by dividing the original image into two parts. Note that both methods still need to use the FFT during processing to accelerate the computation. Compared with conventional FPM reconstruction, both two methods are embedded into the FPM algorithm and can remove the artifacts, but DCT-FPM needs more computation time and computer memory, while PPSID-FPM has a comparable efficiency to conventional FPM. Therefore, the PPSID method is illustrated in detail and verified by both simulations and experiments. Verified by the standard resolution target, the PPSID-FPM algorithm improves the standard deviation of phase accuracy as a factor of 4 from 0.08 radians to 0.02 radians. Finally, we summarize all the existing system errors of FPM within a generalized model. To the best of our knowledge, this may be the last piece of fragment to achieve high-precision FPM.

## 2 Theory and methods

The FPM setup and procedure can be referred to previous work [1-3] and will not be introduced in detail. A typical LR image in FPM is given by

$$I_i(r) = \left| \mathcal{F}_{[O(k-k_i)P(k)]}(r) \right|^2 \tag{1}$$

where $r$=($x$, $y$) denotes the lateral coordinates in the sample plane, $k$=($k_x$, $k_y$) denotes the lateral coordinates in the Fourier domain, $k_i$= ($k_{x,i}$, $k_{y,i}$) is the spatial frequency of the local plane wave emitted by each LED; $P(k)$ is the pupil function, $O(k\text{-}k_i)$ represents the exit wave at the pupil plane, and subscript $i$ is the sequence number of LR images; $\mathscr{F}\{\cdot\}$ is the 2D Fourier transform, which can be defined by

$$\mathcal{F}_{[g]}\left(x,y\right)=\iint g\left(k_x,k_y\right)\exp\left[-j\left(xk_x+yk_x\right)\right]dk_x dk_y \tag{2}$$

where $g(\cdot)$ is a virtual function, $j$ is the imaginary unit, and the original 2π in the Fourier transform is contained in the $k$ space. FFT can calculate the discrete Fourier transform (DFT) of an image at fast speed. However, an implicit periodization assumption of the image is required in FFT [33, 34]. If the image is aperiodic, the assumption may cause a cross-shaped artifact in the Fourier spectrum, termed edge effect as illustrated in Fig. 1. An 11×11 LED array (4 *mm* spacing, central wavelength 630 *nm*) is used in simulations. The distance between the LED array and the sample is 76 *mm*. A dataset of a tile (128×128 pixels) is captured by a 4×/0.1 NA objective and a camera with a 6.5 *μm* pixel pitch. The ground truth of intensity, phase, and its Fourier spectrum in numerical simulations are shown in Figs. 1(a1-c1), respectively, where there are no cross-shaped artifacts in the Fourier spectrum. All the results run 30 iterations to ensure convergence. After FPM reconstruction with the up-sampling of LR image of normal incidence as initial guess, there are cross-shaped artifacts as shown Figs. 1(c2), and both the intensity and phase reconstructions have accuracy error (red circles and arrows). The right pseudo-color is orange and yellow in intensity and phase, respectively, while it is yellow and cyan in reconstructions, respectively. A common and conventional method is to filter the high-frequency error out with a bandpass filter (Figs. 1(a3-c3)). However, there is still accuracy error and not much change visually and the conventional bandpass filter is invalid compared Figs. 1(b2) and (b3). Though the high-frequency artifacts are blocked in the Fourier spectrum, those low-frequency artifacts remain (red arrow in Fig. 1(c3)), resulting in such ripple artifacts on the distribution of intensity and phase (Figs. 1(a3, b3)). And the differences between Figs. 1(a2) and (a3) are very tiny (Figs. 1(a4)). The intensity only has minute differences at the edge (Fig. 1(a4)). Generally, the root-mean-square error (RMSE) is used to evaluate the reconstructions of FPM, which is given by

$$RMSE=\sqrt{\frac{\sum_{x=1}^{M}\sum_{y=1}^{N}|f(x,y)-g(x,y)|^2}{M\cdot N}} \tag{3}$$

where $f(x, y)$ and $g(x, y)$ are two virtual images, $M$ and $N$ denote the size of images. The RMSE of the recovered intensity and phase before and after using the bandpass filter is 4.36% and 4.34% respectively, and 2.47% and 2.47%, respectively, which also indicates the above conclusion.

More terribly, the imaging precision of the same subregion depends on the different sizes of block processing due to different edge conditions, which is illustrated in Figs. 1(a5-c5). Compared Fig. 1(b1), Fig. 1(b3) and Fig. 1(b5), the right pseudo-color is yellow in phase, while it is cyan or orange in reconstructions for different block processing, respectively. And obvious fluctuations are shown at the edge of Fig. 1(b5), which are caused by the high frequency artifacts (red arrows in Fig. 1(c5)) within the passband. In fact, there are also similar artifacts in Fig. 1(c3), but it is not so obvious, because the artifacts are covered by the Fourier spectrum of samples in Fourier space and are also related to the size of aperture synthesis.

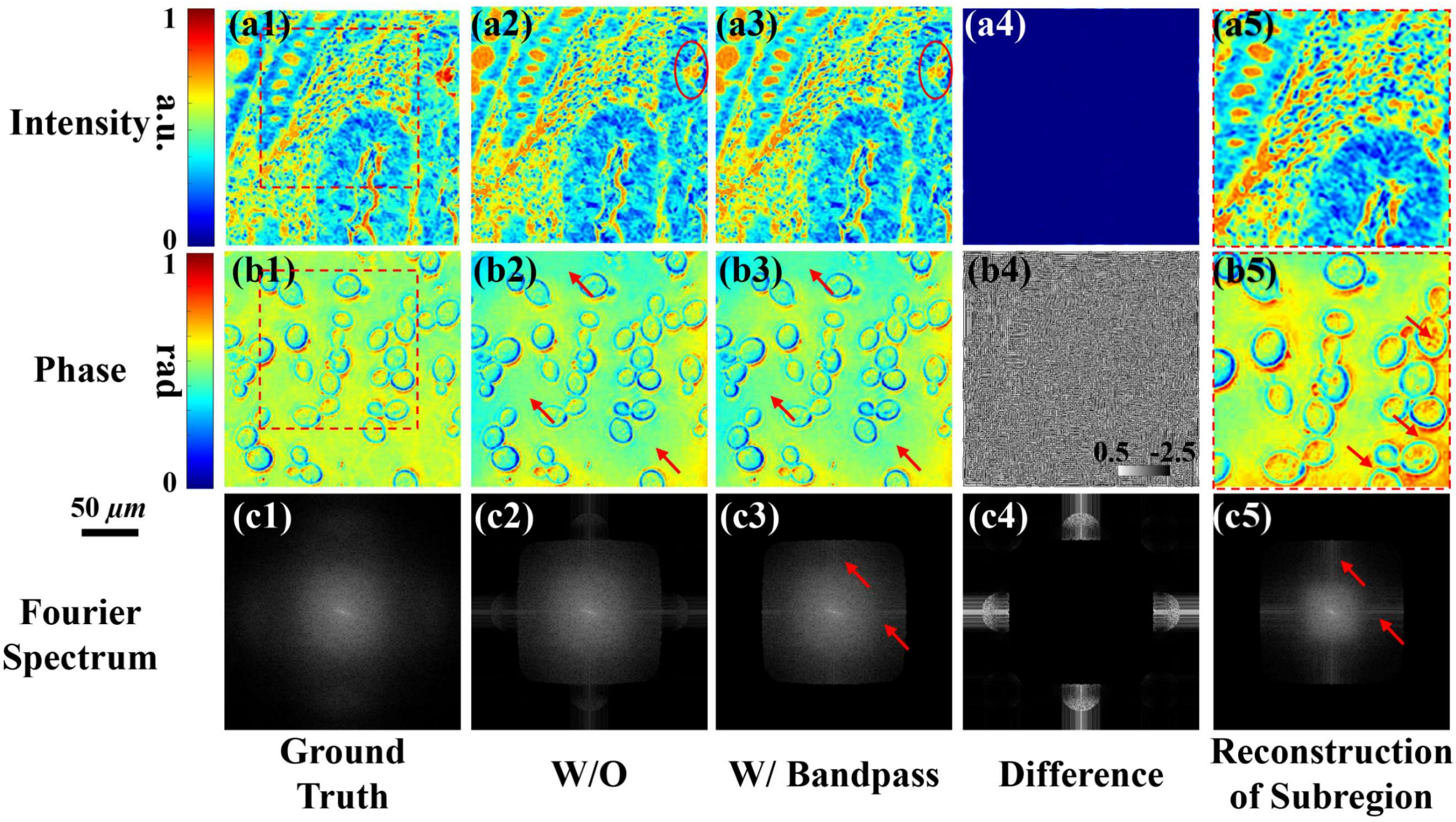

**Fig. 1** Illustration of edge effect. (a1-c1) Ground truth of intensity, phase, and its Fourier spectrum in simulations. Note that there is no cross-shaped artifact in Fourier spectrum. (a2-c2) and (a3-c3) Recovery without and with the conventional bandpass filter, respectively. The artifact cannot be removed by conventional bandpass filter (Red arrows and circles). (a4-c4) Differences between (a2-c2) and (a3-c3). (a5-c5) Reconstruction of a subregion (red box in Fig. 1(a1) and Fig. 1(b1)). Compared Figs. 1(a3, b3) and (a5, b5), the imaging precision depends on the different block processing and fluctuations are obvious at the edge of FOV.

One intuitive hypothesis is that these low-frequency artifacts are caused by the initial guess, therefore, we changed different up-sampling methods as for the initial guess as shown in Fig. 2. By comparing the Fourier spectrum of different initial guesses and reconstructions, it turns out that bilinear guess and bicubic guess cause high-frequency artifacts (red arrows in Figs. 2(d1, d2)) due to up-sampling. However, low-frequency artifacts remain in the reconstructed Fourier spectrum of both ones guess and random guess (red arrows in Figs. 2(d3, d4)), even though no high-frequency artifact is induced by up-sampling. The RMSE of the recovered intensity images of bilinear guess, bicubic guess, ones guess, random guess are 4.36%, 4.35%, 4.55%, 30.75%, respectively. And the RMSE of the recovered phase images are 2.47%, 2.47%, 2.40%, 4.64%, respectively. It turns out that random guess is more difficult to converge (Fig. 2(c4)), compared with the other three initial guesses methods.

Combined with Fig.1, therefore, the accuracy error (Figs. 2(c1-c3)) is mainly caused by the low-frequency artifacts, which are not related to the initial guesses and up-sampling methods. Different initial guesses will result in similar results except for random guess. Herein, we keep using the ones guess in the following simulations and experiments to avoid those high-frequency artifacts, while readers can also choose the conventional bilinear guess or bicubic guess with a bandpass filter to remove those high-frequency artifacts.

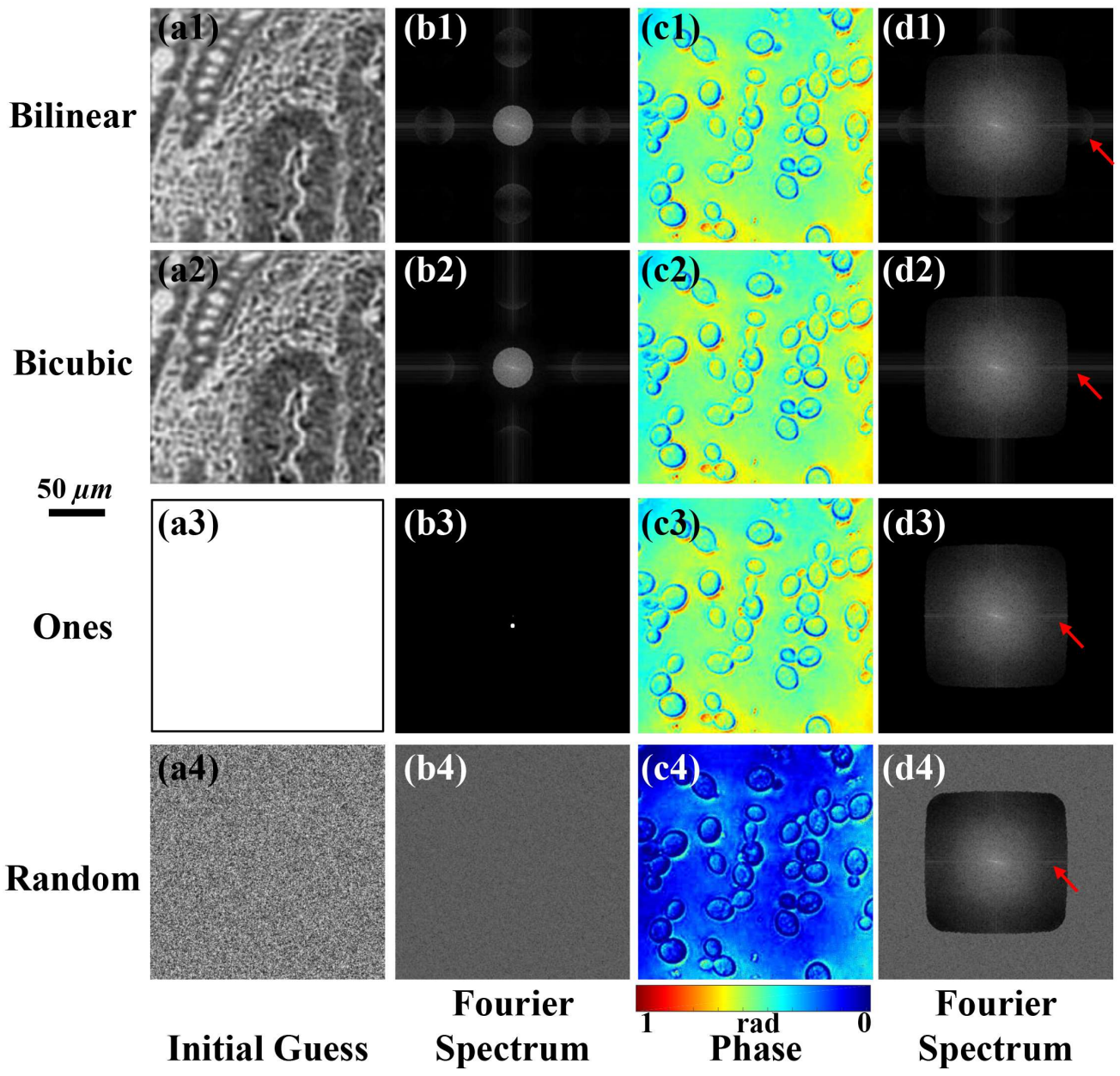

**Fig. 2** Reconstructions of different initial guesses. (a1-a4) Raw images of different initial guesses. (b1-b4) Fourier spectrum of initial guess, respectively. (c1-c4) Phase recovery, respectively. (d1-d4) Fourier spectrum of reconstructions, respectively.

Figure 3 illustrates the cause of edge effect, i.e., the implicit periodization assumption, and the principles of reported solutions, DCT and PPSID. FFT is necessary to quickly calculate the Fourier spectrum of a digital image. However, when implementing FFT, digital images have to be periodically broadened. The aperiodic images are imprecisely regarded as periodic images (Fig. 3(a)), resulting in the cross-shaped artifacts, termed edge effect. To tackle this problem, there are two kinds of opposite methods. One is to add some information to ensure the periodic requirement. Typically, DCT can be regarded as a process that symmetry operation is successively performed along both horizontal and vertical axis to obtain a new image which has four times the size of the original image (Fig. 3(b)) and then FFT is applied to the new symmetrical image for quick calculation. The new image is spatial periodic. Thus, the edge effect is removed. The final image will be one-fourth of the symmetrical image and a cutting operation has to be utilized. So the DCT requires more computation time and computer memory.

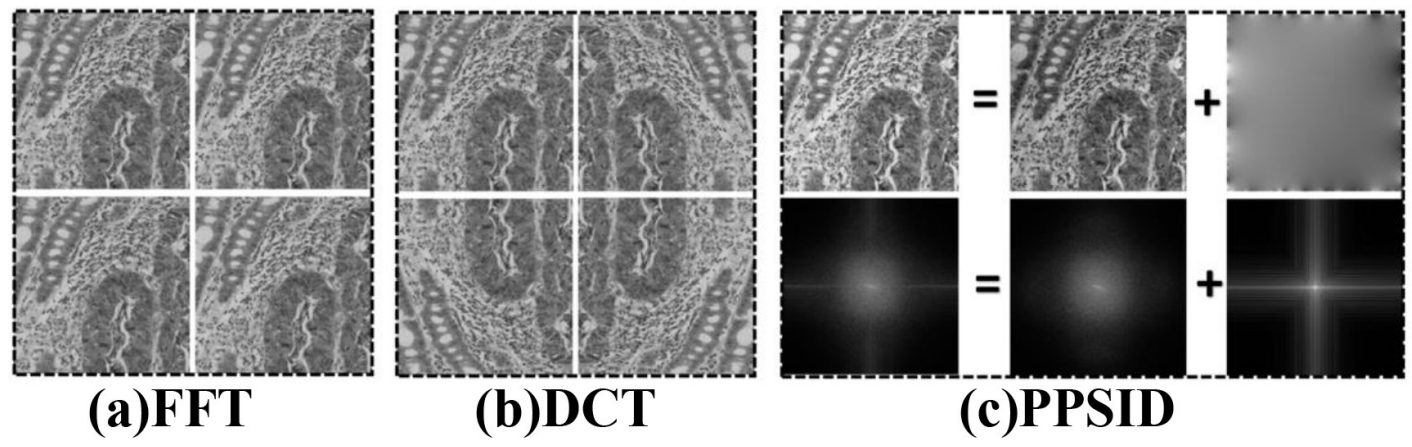

**Fig. 3** Principle of FFT, DCT, and PPSID. (a) Implicit periodization assumption of FFT. (b) Symmetry operation along both horizontal and vertical axis in DCT. (c) Decomposition of the recovered image into a high-precision image and a smooth error image in PPSID.

On the opposite, another kind of methods is to directly remove the error after each loop of FPM reconstructions. Specifically, PPSID approach decomposes an M×N recovered image into two parts (Fig. 3(c)), high-precision results and a smooth error image, which is corresponding to the cross-shaped artifacts. The smooth image will be abandoned in FPM reconstructions with PPSID approach. Numerically, Let $x \in$ [0, M-1] and $y \in$ [0, N-1]. The Fourier transform of the recovered image is extracted by PPSID as follows:

$$\mathcal{F}[g(x,y)] = \mathcal{F}[f(x,y)] - \mathcal{F}[e(x,y)] \tag{4}$$

where $f(x, y)$ denotes the initial image, $g(x, y)$ is the high-precision image we want, and $e(x, y)$ is the smooth error image. To calculate $e(x, y)$, $Q$ is set as the spatial domain of the initial discrete image $f(x, y)$, symbol $a$ is set as a certain pixel in $Q$ space as illustrated in Fig. 4, Ψ space is the expansion of $Q$ space by M×N times when doing FFT, and the symbol $\Psi \backslash Q$ denotes the part of Ψ space except for $Q$, G(a) is defined as the adjacent pixels of pixel $a$ in $Q$ space, and the symbol $b$ denotes the elements of G(a), which is given by:

$$G(a) = \{b \in Q, | a - b | = 1\} \tag{5}$$

H(a) is defined as the adjacent pixels of pixel $a$ in $\Psi \backslash Q$ and the symbol $z$ denotes the elements of H(a), which is given by:

$$H(a) = \{ \exists z \in \Psi \backslash Q \text{ and } | a - z | = 1\} \tag{6}$$

And the symbol $z'$ is defined as the position of pixel $z$ mapping to $Q$ space.

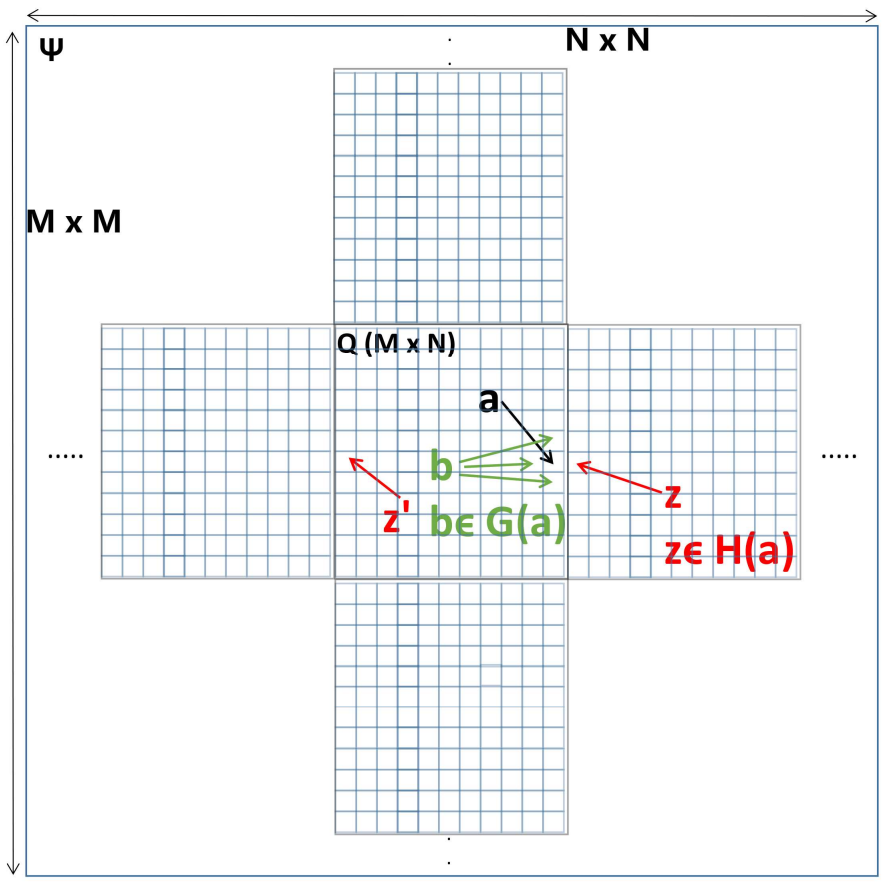

**Fig.4** Illustration of variables and sets in the derivation of PPSID.

To decompose the initial image into two components, there are three basic rules. First, the boundaries of $g(x, y)$ should be as smooth as possible to meet the implicit periodization assumption. Second, the error image $e(x, y)$ should be smooth, since it is corresponding to the low-frequency artifacts. Third, the average intensity of $e(x, y)$ should be zero, the intensity of $g(x, y)$ should not change compared with $f(x, y)$.Thus, the two components $g(x, y)$ and $e(x, y)$ should minimize the object function $E$:

$$\min_{g,e} E = \sum_{\substack{a\in Q, z\in\Psi\backslash Q,\\ |a-z|=1}} (g(a)-g(z'))^2 + \sum_{\substack{a\in Q, b\in Q,\\ |a-b|=1}} (e(a)-e(b))^2 + [\frac{1}{|Q|}\sum_{a\in Q} e(a)]^2 \tag{7}$$

under the constraints:

$$\begin{aligned} & f(x,y) = g(x,y) + e(x,y) \\ & \frac{1}{|Q|}\sum_{a\in Q} e(a) = 0 \end{aligned} \tag{8}$$

Combined with the above constraints, the object function Eq. (7) becomes

$$\min_{e} E = \sum_{a\in Q}( \sum_{z\in H(a)} (f(a)-e(a)-f(z')+e(z'))^2 + \sum_{b\in G(a)} (e(a)-e(b))^2) \tag{9}$$

When $E$ reaches the minimum, the derivative of $E$ with respect to e(a) should be zero, that is:

$$4\sum_{z\in H(a)} (-f(a)+e(a)+f(z')-e(z'))+4\sum_{b\in G(a)} (e(a)\text{-}e(b)) = 0 \tag{10}$$

and then we have

$$\sum_{z\in H(a)} (f(z')-f(a)) = -(|\mathrm{G}(a)|+|H(a)|)\mathrm{e}(a) + \sum_{b\in G(a)} e(b) + \sum_{z\in H(a)} e(z') \tag{11}$$

The absolute value of a set represents the cardinality of the set. Note that

$$|G(a)|+|H(a)|=4 \tag{12}$$

we have

$$L(f) = K * e \tag{13}$$

where * represents the convolution operator, $K$ is defined by

$$K(c) = \begin{cases} -4 & if \;\; c=0 \\ 1 & if \;\; |c|=1 \\ 0 & else \end{cases} \tag{14}$$

where the symbol $c$ is the distance from a certain pixel to pixel $a$ in $\Psi$ space and $L(f)$ is a linear operator defined by

$$L(f)(a)=\sum_{z\in H(a)}(f(z')-f(a)) \tag{15}$$

One easily checks that

$$L(f)=u \tag{16}$$

where ∀ $(p, q) \in$ Q, $p \in [0, M\text{-}1]$ and $q \in [0, N\text{-}1]$, $u(x, y)$ is the four edges of the initial image. Except for the four sides, all other values in $u(x, y)$ are zeros. The value of each side of $u(x, y)$ is equal to the side opposite to it in $f(x, y)$ minus the side corresponding to it in $f(x, y)$. The values of upper and lower sides and values of left and right sides are defined as $u_1$ and $u_2$, respectively. Thus $u=u_1+u_2$, which are given by:

$$\begin{aligned} u_1&=\begin{cases} f(\mathrm{M}-1-p,q)-f(p,q) & if\ p=0\ or\ p=M-1 \\ 0 & else \end{cases} \\ u_2&=\begin{cases} f(p,\mathrm{N}-1-q)-f(p,q) & if\ q=0\ or\ q=\mathrm{N}-1 \\ 0 & else \end{cases} \end{aligned} \tag{17}$$

and that ∀ $(x, y) \in Q$,

$$\mathcal{F}[K(x,y)]=2\cos(\frac{2\pi x}{M})+2\cos(\frac{2\pi y}{N})-4 \tag{18}$$

By taking FFT on both sides of Eq. (13), the smooth error component $e(x, y)$ can be calculated by:

$$\mathcal{F}[e(x,y)]=\begin{cases} \dfrac{\mathcal{F}[u(x,y)]}{2\cos(\frac{2\pi x}{M})+2\cos(\frac{2\pi y}{N})-4} & if\,(x,y)\neq(0,0) \\ 0 & if\,(x,y)=(0,0) \end{cases} \tag{19}$$

and then according to Eq. (4), we obtain the high-precision component $g(x, y)$.

## 3 Simulations and experimental results

The flowchart of different algorithms is compared in Fig. 5 and both the DCT and PPSID methods are embedded into the FPM algorithm. The reconstructions with conventional FPM, DCT-FPM, and PPSID-FPM are shown in Fig. 6, respectively. The accuracy error is highlighted by red circles and arrows. The RMSE of the reconstructed intensity image of conventional FPM, DCT-FPM, and PPSID-FPM is 4.55%, 4.49%, and 0.64%, respectively, and the RMSE of phase image is 2.4%, 1.1%, and 1.1%, respectively. Both DCT-FPM and PPSID-FPM can remove the low-frequency artifacts (Figs. 6(c3, c4)), and improve the accuracy of the phase. The RMSE of phase reduces by half and the ripples disappear. However, the RMSE of intensity has a small drop compared to conventional FPM and DCT-FPM, the accuracy of intensity is not improved (red circle in Fig. 6(a3)). And there is an inconspicuous black solid box in the Fourier spectrum (blue arrow in Fig. 6(c3)). Though we are not very sure the reason of this phenomenon and intensity sometimes is not so important than phase, both intensity and phase image of PPSID-FPM are highly close to ground truth (Figs. 6(a4-c4)). In addition, the time cost of PPSID-FPM is 5.52 s which is comparable to that of conventional FPM, 5.44 s, while the time cost by DCT-FPM is 12.45 s, which is more than twice the others. Note that generally the computation time is not linear to the amount of pixel. Therefore, DCT-FPM can reduce the accuracy error of the phase image and the cross-shaped artifacts in the Fourier spectrum to a certain extent, but it has a small effect on the accuracy error of the intensity image and has a higher time cost, while PPSID-FPM is a fast method of implementing Fourier transform without edge effect.

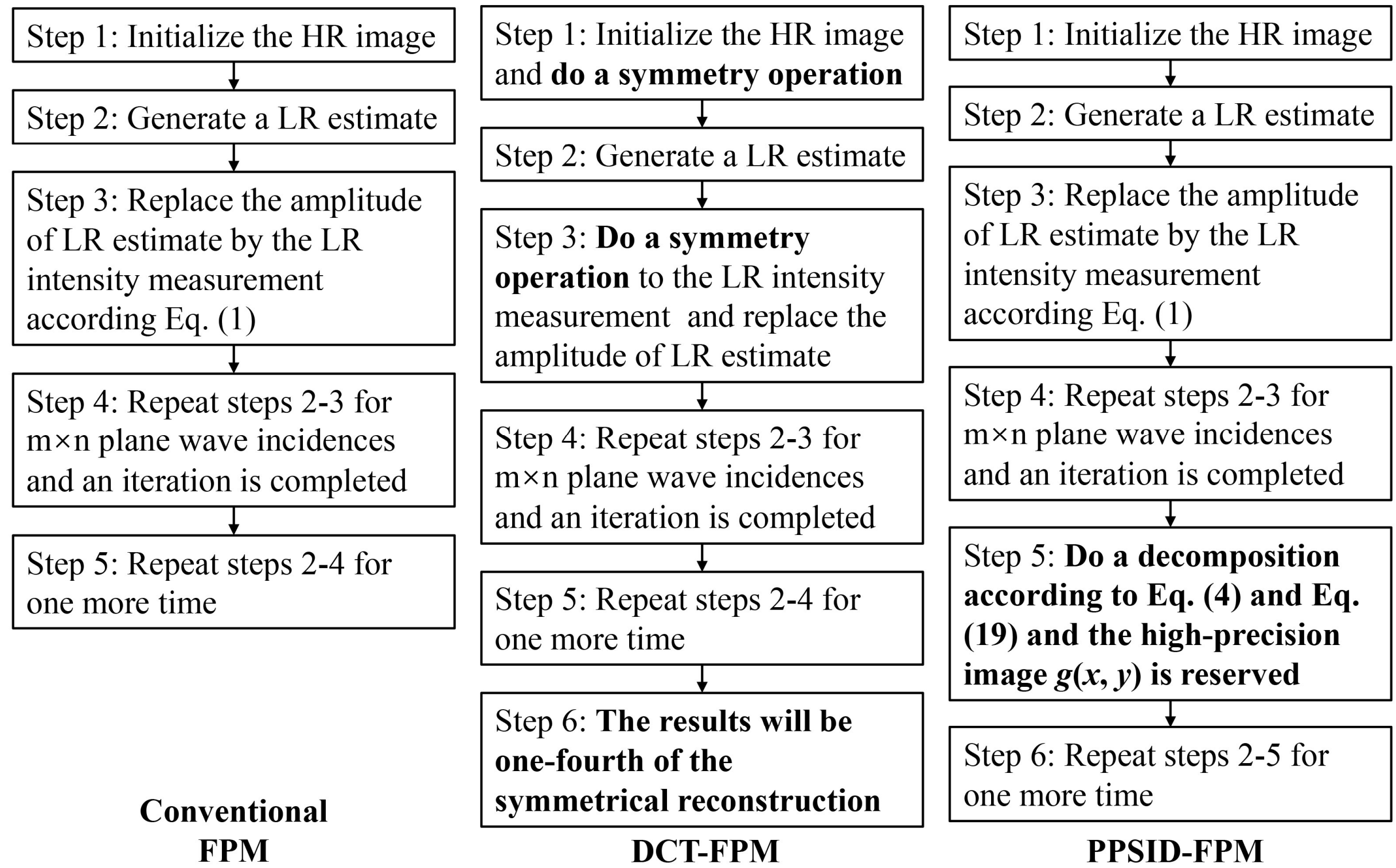

**Fig. 5** The comparison of flowchart between different algorithms in FPM.

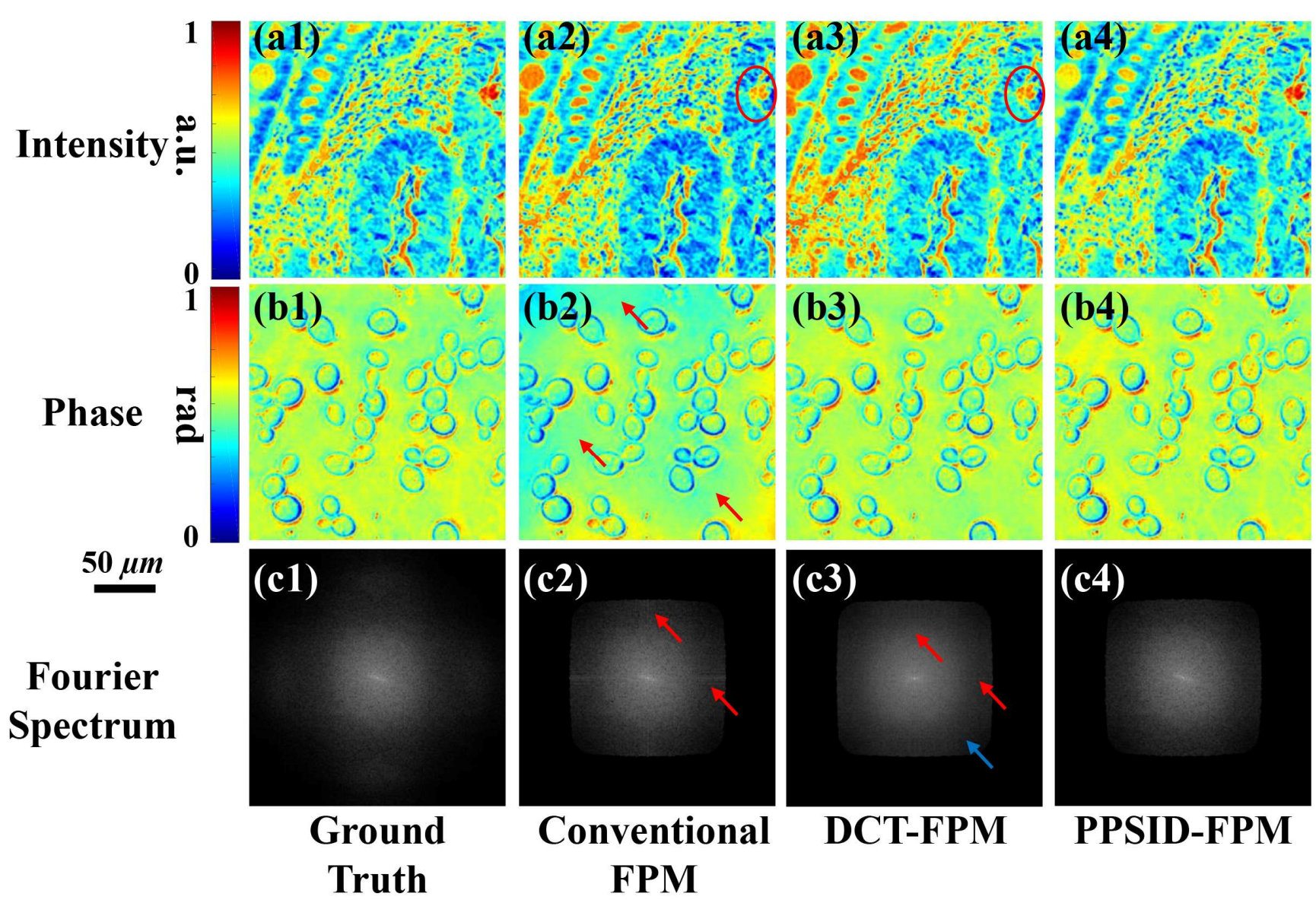

**Fig. 6** Reconstructions of FPM with conventional FPM, DCT-FPM, and PPSID-FPM, respectively. (a1-c1) Ground truth. (a2-c2) Recovery with conventional FPM. (a3-c3) Recovery with DCT-FPM. (a4-c4) Recovery with PPSID-FPM.

In our experiment, a 32×32 programmable LED array (Adafruit, 4 *mm* spacing, central wavelength 631 *nm*, controlled by an Arduino) is utilized, while 15×15 center LED elements are lighted up for imaging and images are captured by a 4×/0.1 NA plan achromatic objective and an 8-bit CCD camera (DMK23G445, Imaging Source Inc., Germany, 1280×960 pixels, 3.75 $\mu m$ pixel each). A resolution target as the sample is placed at 86 *mm* far from the LED array. The entire FOV of the resolution target is shown in Fig. 7(a) and its close-up is shown in Fig. 7(a1), conventional FPM and PPSID-FPM algorithms are applied in FPM Gauss-Newton reconstruction [32], respectively. Currently, we have already obtained quite good results as shown in Fig. 7(b) and other our works at least from the perspective of visual inspection, however, if carefully enlarge the figures and check the background, we can find some imperceptible ripples. The reconstructions with PPSID-FPM algorithm and the difference between conventional FPM and PPSID-FPM reconstructions are given in Figs. 7(c) and 7(d), respectively. It would be much obvious as shown in the difference of phase. A line profile of phase at the background is provided for quantitative measurement as shown in Fig. 7(e). The ground truth of the background is around zero as shown in the black dash line of Fig. 7(e). The standard deviation of phase accuracy with conventional FPM method is 0.08 radians, while it is 0.02 radians with PPSID-FPM algorithm, therefore, the PPSID-FPM algorithm improves the standard deviation of phase accuracy as a factor of 4 from 0.08 radians to 0.02 radians. In terms of computation time, the time cost of PPSID-FPM is 12.33 s, which is comparable to that of conventional FPM of 11.92 s with the CPU of I7-10700 and Matlab 2016a.

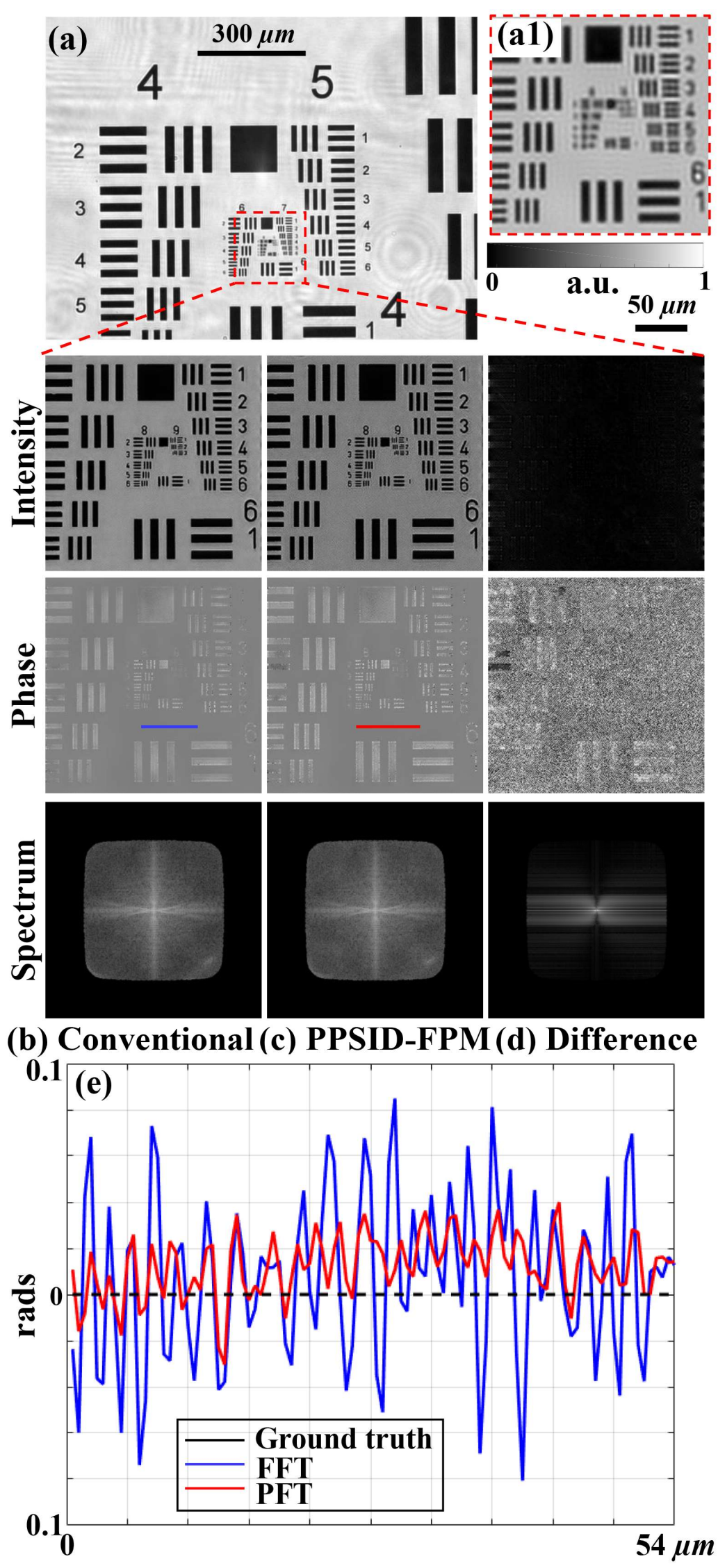

**Fig. 7** Reconstructions of a USAF resolution target. (a, a1) Entire FOV and its close-up. (b) Recovery with conventional FPM approach. (c) Recovery with PPSID-FPM method. (d) Differences between these two methods. (e) Line profile of phase accuracy.

To further verify the effectiveness of PPSID-FPM algorithm, we use a Siemen Star and a Hela cell section as the samples, and it is not stained so that it can be regarded as a pure phase object as shown Fig. 8. The experiment with the Siemen Star and the Hela cell was captured by a 20×/0.4 NA with the same camera in order to test the PPSID-FPM performance under different objective and a circle illumination pattern was used with the NA of illumination of 0.41. Considering less flux of this objective, we change the height to 55 *mm* to get enough brightness. The missing part of the spectrum is because of the vignetting effect and some unperfect images are omitted [30]. The results of first two rows are from the Siemen Star and the rest are from the Hela cell. The ripples would be much obvious for the pure phase samples compared the Fig. 8(a) and Fig. 8(b). In terms of computation time, the time cost of PPSID-FPM for Siemen Star is 16.62 s, which is comparable to that of conventional FPM of 15.94 s. The time cost of PPSID-FPM for Hela cell is 8.51 s, which is comparable to that of conventional FPM of 7.97 s.

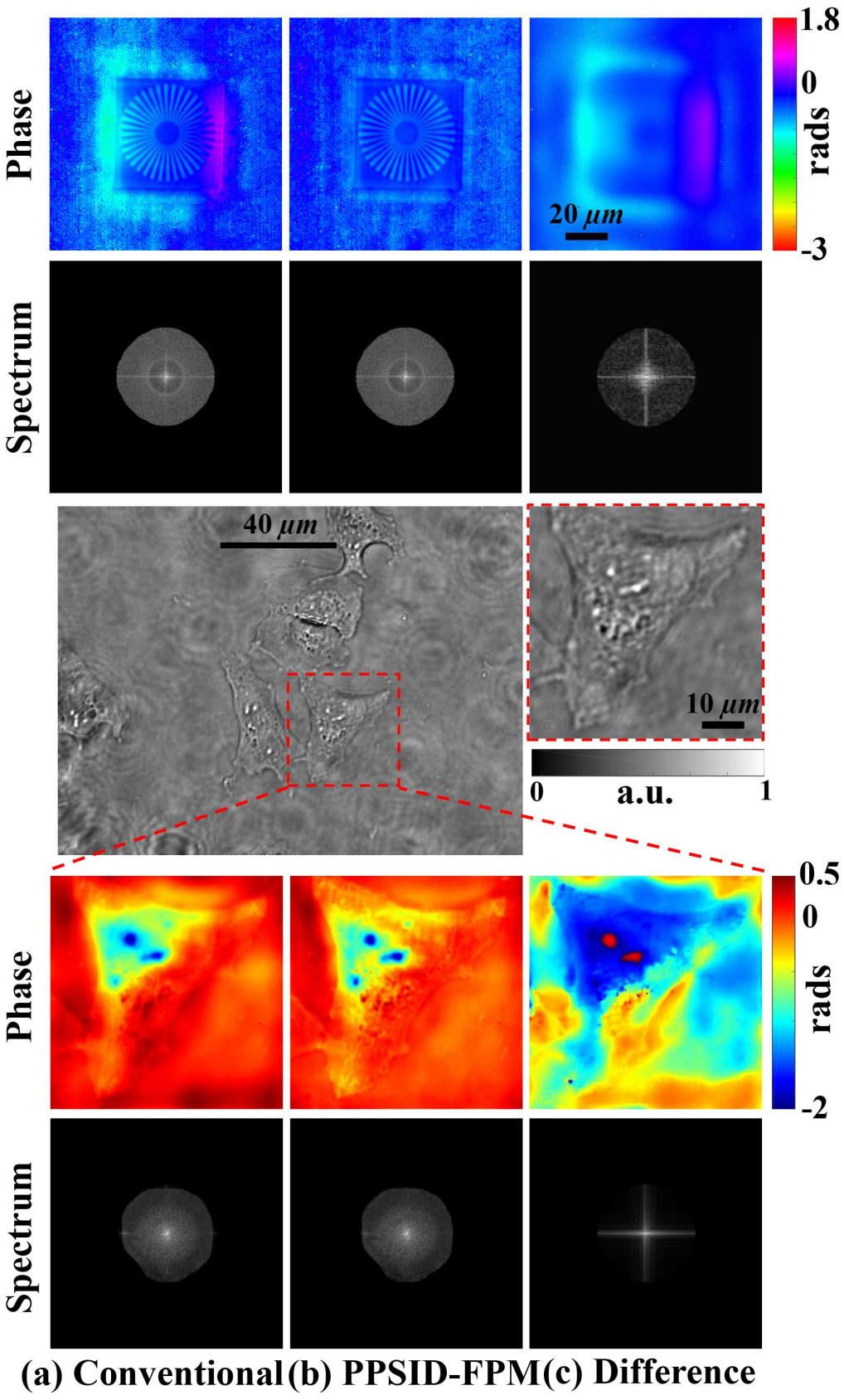

**Fig. 8** Reconstructions of a phase only Simen Star and a Hela cell section in experiments. (a) Recovery with conventional FFPM approach. (b) Recovery with PPSID-FPM method. (c) Differences between these two methods.

## 4 Conclusions and discussions

In conclusion, the artifacts caused by the edge effect would degrade the imaging precision of FPM, especially the phase recovery, and it depends on the different sizes of block processing due to different edge conditions. The edge effect is the conflict between the aperiodicity of specimens and the implicit periodization assumption of FFT, which makes the zero-padding interpolation failed. This artifact is a cross shape in Fourier space spanning from high frequency to low frequency and is not relevant to initial guess. The high frequency parts are corresponding to the fluctuations at the edge of FOV in spatial domain and can be removed by the conventional bandpass filter. The low frequency parts look like ripples at the background of reconstructions. Note that the artifact is not always a perfect cross shape in Fourier space as shown in Fig. 7(d) and is related to the edge conditions. The visibility of artifacts is related to the Fourier spectrum of object and the size of aperture synthesis. Since this artifact is ranging from high frequency to low frequency and high frequency part would be easier to be perceived. Therefore, if the size of aperture synthesis is not very large, the artifacts may be covered by the object's spectrum. Meanwhile, it would be much obvious for phase targets. And the artifacts only effect the edge of the recovery of intensity, while it has greater impact on phase accuracy. We reported two kinds of opposite methods, termed DCT and PPSID to remove the edge effect in FPM, and PPSID-FPM has lower time cost and better performances in terms of recovered intensity. Therefore, the PPSID-FPM method is illustrated and verified in detail. The PPSID-FPM approach is to subtract the low frequency error from the results of convention FPM, which can be regarded as decomposing one aperiodic image into a periodic image and a smooth error image. The efficiency of PPSID-FPM is comparable to conventional FPM. It would have a significant application in quantitative biology and measurements. We also provide the open demo codes for uncommercial use [37].

To the best of our knowledge, this work may be the last piece of fragment in terms of the system errors in FPM. Combined with the other known system errors in FPM, the model of Eq. (1) can be modified to

$$I_i(r) = w_i \left| \mathcal{F}_{\left[O(k-k_i-\Delta k_i)P(k)e^{j\phi}\right]}(r) \right|^2 + I_{n,i} \tag{20}$$

where the weight factor $w_i$ is introduced by the LED intensity fluctuations, the phase term $e^{j\Phi}$ is introduced by the aberrations, which is included in the pupil function, the intensity term $I_{n,i}$ is introduced by the noise. And the positional misalignment would make the estimation of $k_i$ not precise and introduce the term $\triangle k_i$. The $\mathcal{F}$ originally denotes the FFT and now will be replaced by the PPSID.

We must admit that the current phase accuracy of FPM is quite high and has been used for some measurements [3] with the accuracy of around λ/100. But higher accuracy is always a constant pursuit. The PPSID-FPM algorithm improves the standard deviation of phase accuracy as a factor of 4 to 0.02 radians, which is around λ/400. Undoubtedly, we have to admit that the phase accuracy recovered by our experiments is not exact completely and need more experiments with different phase targets. Though this work does not aim to provide the exact measurement of accuracy, it indicates the potential to improve the accuracy. And the exact measurement of accuracy would be our future work.

*Disclosures*

The authors have no relevant financial interests in this article and no potential conflicts of interest to disclose.

*Acknowledgments*

We acknowledge the funds from National Natural Science Foundation of China (NSFC) (12104500). An Pan thanks Daniel Martin (California Institute of Technology, USA) for helpful discussions and inspiration for this work.